\documentclass[12pt]{article}

\catcode`@=11 \@addtoreset{equation}{section} \catcode `@=12

\begin{document}

\begin{center}
{\large\bf On the role of Planck's oscillator in the construction of
Heisenberg's mechanics} \\[.5cm]
{\large Budh Ram} \\
Physics Department, New Mexico State University \\
Las Cruces, New Mexico 88003, USA \\[.1cm]
and \\[.1cm]
Prabhu-Umrao Institute of Fundamental Research \\
A2/214 Janak Puri, New Delhi 110 058, India
\end{center}

\vspace*{.5cm}

In the present note we elucidate how physical considerations
based on Planck's oscillator led to the construction of Heisenberg's
mechanics.

\vspace*{.5cm}

I read with the greatest interest the recent paper$^1$ by Aitchison,
MacManus and Snyder on the epoch-making July 1925 paper$^2$ of
Heisenberg.  They say that this paper of Heisenberg is widely regarded
as being difficult to understand mainly because he gave remarkably few
details of the calculations he performed; and in particular, he simply
wrote down the quantum version:
$$
\cos\omega(n,n-0)t: \omega^2_0 a_0(n) + {1\over4} \left[a^2(n+1,n) +
a^2(n,n-1)\right] = 0,
\eqno (1a)
$$
$$
\cos\omega(n,n-1)t: -\omega^2(n,n-1) + \omega^2_0 = 0, \hspace*{1.55in}
\eqno (1b)
$$
$$
\cos\omega(n,n-2)t: \left[-\omega^2(n,n-2) + \omega^2_0\right]
a(n,n-2) \ + \hspace*{1.1in}
$$
$$
\hspace*{1.3cm} {1\over2}\left[a(n,n-1) a(n-1,n-2)\right] = 0,
\eqno (1c)
$$
$$
\cos\omega(n,n-3)t: \left[-\omega^2(n,n-3)+\omega^2_0\right] a(n,n-3) \
+ \hspace*{1in}
$$
$$
{1\over2} \left[a(n,n-1) a(n-1,n-3) + a(n,n-2)
a(n-2,n-3)\right] = 0,
\eqno (1d)
$$
of the classical equations:
$$
\cos 0\omega t: \omega^2_0 a_0 (n) + {1\over2} a^2_1 (n) = 0, \hspace*{.5in}
\eqno (2a)
$$
$$
\cos 1\omega t: (-\omega^2 + \omega^2_0) = 0, \hspace*{.9in}
\eqno (2b)
$$
$$
\cos 2\omega t: \left(-(2\omega)^2 + \omega^2_0\right) a_2 + {1\over2}
a^2_1 = 0,
\eqno (2c)
$$
$$
\cos 3\omega t: \left(-(3\omega)^2 + \omega^2_0\right)a_3 + a_1 a_2 =
0,
\eqno (2d)
$$
which are obtained by substituting
$$
x = \lambda a_0 + a_1 \cos \omega t + \lambda a_2 \cos 2\omega t +
\lambda^2 a_3 \cos 3\omega t + \cdots + \lambda^{\tau - 1} a_\tau \cos \tau
\omega t
\eqno (3)
$$
in the classical equation of motion
$$
\ddot{x} + \omega^2_0 x + \lambda x^2 = 0
\eqno (4)
$$
for the anharmonic oscillator, the left-hand sides (LHS) of Eqs. (2)
being, respectively, the coefficients of $\cos 0\omega t$ $(= 1)$,
$\cos 1\omega t$, $\cos 2\omega t$ and $\cos 3\omega t$, which we have
expressly indicated.  (The reason for doing so will become evident
soon.) 

How did Heisenberg arrive at Eqs. (1)?, ask Aitchison {\it et
al}$^1$.  And then they go on to show how Eqs. (1) are derived from
the recursion relation, Eq. (22) of their paper$^1$, and conjecture
that this is what Heisenberg$^2$ did, calling their approach to be
Heisenberg's. 

In the present note we elucidate how physical considerations$^4$
based on the quantum oscillator of Planck enabled Heisenberg to simply
write down Eqs. (1) -- with no details necessary.

The energies of the stationary states of the Planck oscillator are
given by 
$$
W(n) = {n h \omega \over 2\pi},
\eqno (5)
$$
with $\omega$ the frequency of the oscillator $x = a_1\cos \omega t$.
Let us label these [Planck stationary states] as
\[
\cdots,n-3,n-2,n-1,n,n+1,n+2,n+3,\cdots .  
\]
Eq. (5) can be rewritten in the form
\begin{eqnarray}
\omega &=& {2\pi \over h} \left[W(n) - W(n-1)\right] = {2\pi \over h}
\left[W(n+1) - W(n)\right], \hspace*{1.1in} (6) \nonumber \\
&& \hspace*{1.5cm} {\buildrel {n \rightarrow n-1} \over {\rm emission}} \hspace*{3cm}
{\buildrel {n+1 \leftarrow n} \over {\rm absorption}} \hspace*{4cm} \nonumber
\end{eqnarray} 
which says that frequency of the emitted or absorbed Planck
quantum (light) is the same$^5$ as that of the oscillator
$(a_1 \cos\omega t)$ and that the Planck oscillator makes transitions
\underbar{only} to neighbouring stationary states by emitting or
absorbing \underbar{one} quantum of energy $\displaystyle{h\omega
\over 2\pi}$ or of frequency $\omega$.  Furthermore, if the Planck
oscillator is to go from the state $n$ to the state $n-2$, it got to
emit \underbar{two} quanta of frequency $\omega$, one to make the
transition from the state $n$ to the state $n-1$, and another to make
the transition from $n-1$ to $n-2$.  Now suppose there is another
Planck oscillator but of frequency $2\omega$ $(a_2 \cos 2\omega t)$
which acts \underbar{independently} of the one with frequency
$\omega$.  Then the transition from $n$ to $n-2$ can also be achieved
by the emission of a single quantum of frequency $2\omega$.  In the
same vein, a transition from $n$ to $n-3$ can occur by the emission of
a single quantum of frequency $3\omega$ $(a_3 \cos 3\omega t)$ or by
the emission of two quanta, one of frequency $2\omega$ from $n$ to
$n-2$ and the other of frequency $1\omega$ from $n-2$ to $n-3$, or by
the emission of three quanta each of frequency $1\omega$; and a
transition from $n$ to $n-2$ can occur by the absorption of a single
quantum of frequency $1\omega$ to go from $n$ to $n+1$ and then the
emission of a single quantum of frequency $3\omega$ to go from $n+1$
to $n-2$; etc.

Now in his general theory of line spectra Bohr's second assumption
states that the radiation absorbed or emitted during a transition
between two stationary states $E'$ and $E^{\prime\prime}$ possesses a
frequency $\nu$ $\left(= {\omega \over 2\pi}\right)$, given by the
relation$^6$
$$
\omega = {2\pi \over h} \left(E' - E^{\prime\prime}\right).
\eqno (7)
$$
Note that \underbar{implicit} in this assumption is the notion of the
Planck oscillator.  If one analogously labels the Bohr stationary
states as
\[
\cdots, n-3, n-2, n-1, n, n+1, n+2, n+3, \cdots .
\]
(with $n$ the reference state) then Eq. (7) can be rewritten as
\begin{eqnarray} 
\hspace*{2cm} 
\omega(n,n-1) &=& {2\pi \over h} {\left[W(n) - W(n-1)\right]};
\hspace*{3.5cm} (8a) \nonumber \\ 
&& \hspace{1.5cm} {\buildrel {n \rightarrow n-1} \over {\rm emission}} \nonumber 
\end{eqnarray}
or
\begin{eqnarray}
\hspace*{2cm}
\omega(n,n+2) &=& {2\pi \over h} \left[W(n) - W(n+2)\right];
\hspace*{3.5cm} (8b) \nonumber \\
&& \hspace*{1.5cm} {\buildrel {n \rightarrow n+2} \over {\rm
absorption}} \nonumber
\end{eqnarray}
or

\newpage

\begin{eqnarray}
\hspace*{2cm}
\omega(n+4,n-1) &=& {2\pi \over h} \left[W(n+4) - W(n-1)\right];
\hspace*{2cm} (8c) \nonumber \\
&& \hspace*{2cm} {\buildrel {n+4 \rightarrow n-1} \over {\rm emission}} \nonumber
\end{eqnarray}
etc., the frequencies $\omega(n,n-1), \omega(n,n+2),
\omega(n+4,n-1)$, etc. being different from one another, represent
independent Planck oscillators$^7$.  In other words, the Planck
oscillator of frequency $\omega(n,n-1)$ cannot make a transition from
$n$ to $n-2$ or from $n$ to $n-3$, etc.  Thus if one wants to make a
quantum mechanics of the spectral lines and their intensities, one
needs to make quantum mechanics of the Planck harmonic oscillator as
Bohr's jumps are \underbar{the} jumps (transitions) of the Planck
oscillators$^7$.  Therefore, first and foremost, Heisenberg tried$^8$
to do just that -- with his expertise in Bohr-Sommerfeld theory$^9$,
dispersion theory$^{10}$, and his knowledge of Born's
prescription$^{11}$ to translate classical quantities into quantum
mechanical ones.  And he succeeded.

Heisenberg's construction is reproduced on page 26 of Ref. 3.  We
recast it as follows:

For the simple harmonic oscillator $(\omega = 1\omega)$
$$
x = a_1 \cos\omega t
\eqno (9)
$$
the action variable
$$
J = \oint pdx = \int^{2\pi/\omega}_0 m {\dot x}^2 dt = \pi m a^2_1
\omega;
\eqno (10)
$$
or
$$
1 = \pi m {\partial \over \partial J} (a^2_1 \omega).
\eqno (11)
$$
Then using Born's prescription$^{11}$, (11) becomes
$$
h = \pi m\left[a^2(n+1,n) \omega(n+1,n) - a^2(n,n-1)
\omega(n,n-1)\right]
\eqno (12)
$$
as the Planck oscillator is characterised by only two distinct
amplitudes, $a(n,n+1)$ for absorption of a Planck quantum of frequency
$\omega(n,n+1) = \omega$, and $a(n,n-1)$ for emission of a quantum of
the same frequency $\omega = \omega(n,n-1)$.  Eq. (12) is then
rewritten as 
$$
h = \pi m \omega\left[a^2(n+1,n) -a^2(n,n-1)\right],
\eqno (13)
$$
which has the solution
$$
a^2(n,n-1) = {nh \over m\pi \omega},
\eqno (14)
$$
with the stipulation $a(0,-1) = 0$, i.e. there is a lowest bound for
the Planck energy levels, namely $n=0$, and there can be no emission
of a quantum from it.

Now one can calculate the energy of the $n$-th Planck state of the
oscillator by substituting (9) in 
$$
W = {1\over2} m \left({\dot x}^2 + \omega^2 x^2\right)
\eqno (15)
$$
and obtaining
$$
W = {1\over2} m\omega^2 a^2_1.
\eqno (16)
$$
But as $a^2_1$ has only two distinct possible values, $a^2(n+1,n)$ and
$a^2(n,n-1)$, Eq. (16) becomes
$$
W(n) = {1\over2} m \omega^2 {1\over2}\left[a^2(n+1,n) +
a^2(n,n-1)\right],
\eqno (17)
$$
or, using (14),
$$
W(n) = \left(n + {1\over2}\right) \omega {h \over 2\pi}.
\eqno (18)
$$
Thus Heisenberg made the quantum mechanics of the harmonic
oscillator$^{12,13}$.

But this was not enough.  In order to solve the problem of atomic
spectral lines and their intensities, one needed to know the relation
between different transition amplitudes, say, for example, between
$a(n,n-2)$ [the transition amplitude from state $n$ to state $n-2$ of
frequency $\omega(n,n-2)$] and $a(n,n-1)$ [the transition amplitude
from state $n$ to $n-1$ of frequency $\omega(n,n-1)$], etc.  So
Heisenberg went back to his favourite$^{14}$ anharmonic oscillator.
And there it [the relation he sought for] was in Eq. (2c).  The
amplitude $a_2$, which represents in Planck's oscillator language, the
transition amplitude from state $n$ to $n-2$ of frequency
$\omega(n,n-2)$ \underbar{as well as} from state $n$ to $n+2$ of
frequency $\omega(n,n+2)$, was expressible as the \underbar{product}
of $a_1$ and $a_1$ (and not as the sum of $a_1$ and $a_1$), $a_1$
being the transition amplitude from state $n$ to $n-1$ or from $n-1$
to $n-2$, etc.$^{15}$  In view of the discussion given earlier,
Eq. (2c) has the obvious translation in terms of Bohr labeling:
$2\omega \rightarrow \omega(n,n-2)$; $a_2 \rightarrow a(n,n-2)$;
$a^2_1 \rightarrow a(n,n-1) a(n-1,n-2)$, and is simply written as
Eq. (1c).  Similarly in Eq. (2d), $3\omega \rightarrow \omega(n,n-3)$;
$a_3 \rightarrow a(n,n-3)$; and $a_1 a_2 \rightarrow {1\over2}
[a(n,n-1) a(n-1,n-3) + a(n,n-2) a(n-2,n-3)]$ as the transition from
state $n$ to state $n-3$ can occur in two possible ways: by a
transition from $n \rightarrow n-1$ by the action of $a_1$ followed by
the transition from $n-1$ to $n-3$ by the action of $a_2$; or by a
transition from $n \rightarrow n-2$ by the action of $a_2$ followed by
the transition from $n-2 \rightarrow n-3$ by the action of $a_1$; and
it is simply written as Eq. (1d).  Eq. (2b) obviously translates into
Eq. (1b) which simply verifies that the transition frequency $\omega
(n,n-1)$ of the Planck oscillator is the \underbar{same} as the
frequency of the classical oscillator.  Now the LHS of Eq. (2a) is the
constant term, i.e. it is multiplied by $\cos 0\omega t$, and thus
represents no transition from the state $n$.  Hence $a_0(n)
\rightarrow a_0(n,n-0) \equiv a_0(n)$; since $a^2_1$ can act on state
$n$ in two ways: $a(n,n+1) a(n+1,n)$ and $a(n,n-1) a(n-1,n)$, $a^2_1
\rightarrow {1\over2} [a^2(n+1,n) + a^2(n,n-1)]$ with $a(n,n+1) =
a(n+1,n)$ and $a(n-1,n) = a(n,n-1)$.  And Eq. (2a) translates into
Eq. (1a).$^{16}$  Aitchison {\it et al.}$^1$ add the terms of order
$\lambda^2$ to Eqs. (2) (see Eqs. (27) of Ref. 1), the term added to
(2d) being 
$$
\cos 3\omega t: \lambda^2(a_1 a_4 + 2a_0 a_3) + \cdots .
\eqno (19)
$$
Let us see how it translates into quantum theoretical labeling.  Since
it multiplies $\cos 3\omega t$, it represents the transition from the
state $n$ to $n-3$ of frequency $\omega(n,n-3)$.  Now by the action of
$a_1 a_4$ this can only be achieved by $a(n,n+1) a(n+1,n-3)$ or by
$a(n,n-4) a(n-4,n-3)$, i.e. $a_1 a_4 \rightarrow {1\over2}
\left[a(n,n+1) a(n+1,n-3) + a(n,n-4) a(n-4,n-3)\right]$; and by the
action of $a_0 a_3$ by ${1\over2} \left[a_0(n) a(n,n-3) + a(n,n-3)
a_0(n-3,n-3)\right]$. So the expression (19) translates into
$$
\lambda^2 \Bigg[{1\over2} \left\{a(n,n+1) a(n+1,n-3) + a(n,n-4)
a(n-4,n-3)\right\} + 
$$
$$
\left\{a_0(n) a(n,n-3) + a(n,n-3) a_0(n-3,n-3)\right\}\Bigg],
\eqno (20)
$$
which is the same that appears in Eq. (46) of their paper$^1$.

Put succinctly, one can indeed write down the quantum version of
Eqs. (2) [or Eqs. (27) of Aitchison {\it et al.}$^1$] simply by
considerations based on the Planck oscillator.

One may note that the energies of the stationary states given by (18)
for the simple harmonic oscillator come from (16) -- the constant term
(no periodic terms) -- strongly suggesting that the energies of the
anharmonic oscillator should likewise be given by constant terms which
result after substituting (3) in 
$$
W = {1\over2} m {\dot x}^2 + {1\over2} m \omega^2_0 x^2 + {1\over3} m
\lambda x^3.
\eqno (21)
$$
To first order in $\lambda$ the only constant term that results (after
using trigonometric identities) is ${1\over2} m \omega^2_0 a^2_1$
giving $W(n)$ by Eq. (18).  We emphasize that all periodic terms
vanish --- a fact that has been verified by Aitchison {\it et
al.}$^1$, showing thereby that the quantum theoretical labeling is
indeed \underbar{internally consistent}.

The energies of the stationary states of the anharmonic oscillator 
$$
{\ddot x} + \omega^2_0 x + \lambda x^3 = 0
\eqno (22)
$$
are likewise calculable by substituting
$$
x = a_1 \cos\omega t + \lambda a_3 \cos 3\omega t + \lambda^2 a_5 \cos
5\omega t + \cdots 
\eqno (23)
$$
in the expression
$$
W = {1\over2} m {\dot x}^2 + {1\over2} m \omega^2_0 x^2 + {1\over4} m
\lambda x^4
\eqno (24)
$$
and finding the constant terms after using trigonometric identities.
They turn out to be (to first order in $\lambda$)
$$
W = {1\over2} m \omega^2_0 a^2_1 + {1\over4} m \lambda {3\over8}
(a^2_1)^2,
\eqno (25)
$$
or
$$
W(n) = \left(n + {1\over2}\right)\omega_0 {h \over 2\pi} + \lambda
{3\over8} {\left(n^2 + n + {1\over2}\right) h^2 \over 4\pi^2
\omega^2_0 m},
\eqno (26)
$$
after the use of (14).  One may \underbar{note} that the anharmonic
oscillator (22) represents the Planck oscillators of transition
frequencies 
$$
\omega(n,n-1) = {2\pi \over h} \left[W(n) - W(n-1)\right]
$$
$$
\hspace*{.6cm} = \omega_0 + \lambda {3\over8} {hn \over \pi \omega^2_0 m}
\eqno (27)
$$
and transition amplitudes$^{17}$
$$
a(n,n-1) = \sqrt{nh \over m \pi \omega(n,n-1)} = \sqrt{nh \over
m\pi\omega_0} \left[1 - \lambda {3\over16} {hn \over \pi\omega^3_0
m}\right].
\eqno (28)
$$

Before closing we \underbar{emphasize} that the quantum method given
by Heisenberg$^2$ could be applied to solving the \underbar{real}
problem of atomic spectral lines and their intensities \underbar{only}
in its reformulated form by Born and Jordan$^{18}$, as demonstrated by
Pauli$^{19}$ in his solution of the Hydrogen spectrum$^{20}$.

Last, we may remark that the paper of Aitchison {\it et al.}$^1$
and the present note teach unequivocally the lesson -- which every
student must learn -- that creative work comes about via many routes,
including (i) mathematical expertise and sophistication, and (ii)
physical insight and considerations.
\bigskip

\begin{center}
ACKNOWLEDGEMENTS
\end{center}
\medskip

I thank Heike Moore and Matthias Burkardt for making translations from
German to English for me, and R.S. Bhalerao for a critical reading of
the manuscript.
\bigskip

\begin{center}
REFERENCES AND FOOTNOTES
\end{center}
\medskip

\begin{enumerate}
\item[{1.}] I.J.R. Aitchison, D.A. MacManus, and T.M. Snyder,
``Understanding Heisenberg's ``magical'' paper of July 1925: A new look
at the calculational details'', Am. J. Phys. {\bf 72}, 1370-1379
(2004).

\item[{2.}] W. Heisenberg, ``Quantum-theoretical
re-interpretation of kinematic and mechanical relations'',
Z. Phys. {\bf 33}, 879-893 (1925); English translation in
Ref. 3, pp. 261-276.

\item[{3.}] \underbar{Sources of Quantum Mechanics}, edited by
B.L. van der Waerden (Dover Publications, New York, 1968).

\item[{4.}] In tune with the character of Heisenberg's physics; in
this connection, see Pauli's reaction, Ref. 3, p. 37.

\item[{5.}] Ref. 3, p. 22.

\item[{6.}] Ref. 3, p. 5.

\item[{7.}] In this connection see also Ref. 3, p. 189.

\item[{8.}] The method of trial and error is indispensable to
research; see, for example, B. Ram, ``Presenting the Planck's relation
$E = n h \nu$'', Am. J. Phys. {\bf 42}, 1092-1094 (1974).

\item[{9.}] Ref. 3, pp. 21-22.

\item[{10.}] H.A. Kramers and W. Heisenberg, ``On the dispersion of
radiation by atoms'', Z. Phys. {\bf 31}, 681-708 (1925); English
translation in Ref. 3, pp. 223-252.

\item[{11.}] M. Born, ``Quantum Mechanics'', Z. Phys. {\bf 26},
379-395 (1924); English translation in Ref. 3, pp. 181-198; see
footnote 3 on p. 182.

\item[{12.}] As envisioned by Planck; see I. Duck and
E.C.G. Sudarshan, \underbar{100 Years} \underbar{of Planck's Quantum} (World
Scientific, Singapore, 2000), pp. 51 and 73.

\item[{13.}] One may note that the extra term ${1\over2} \omega {h
\over 2\pi}$ had already appeared in the work of Planck; see
M. Planck, \underbar{The Theory of Heat Radiation} (Dover
Publications, New York, 1991) [originally published in 1914], p. 142. 

\item[{14.}] See Ref. 3, pp. 23-24.

\item[{15.}] Remembering that the Planck oscillator makes transitions 
only to neighboring states.

\item[{16.}] Here we may comment that actually $a^2_1$ in Eq. (16)
should be written as ${1\over2} [a(n,n+1) a(n+1,n) + a(n,n-1)
a(n-1,n)]$, which is equivalent to the expression on the RHS of
Eq. (17). 

\item[{17.}] This verifies that the energies of stationary states must
be given by constant terms.

\item[{18.}] M. Born and P. Jordan, ``On quantum mechanics'',
Z. Phys. {\bf 34}, 858-888 (1925); English translation in Ref. 3,
pp. 277-306.

\item[{19.}] W. Pauli, ``On the Hydrogen spectrum from the stand point
of the new quantum mechanics'', Z. Phys. {\bf 36}, 336-363
(1926); English translation in Ref. 3, pp. 387-415.

\item[{20.}] It can now be rigorously shown that the Hydrogen atom is
equivalent to a four-dimensional harmonic oscillator; see Refs. 21-23.

\item[{21.}] D.S. Bateman, C. Boyd and B. Dutta-Roy, ``The mapping of
the Coulomb problem into the oscillator'', Am. J. Phys. {\bf 60},
833-836 (1992).

\item[{22.}] H.A. Mavromatis, ``Mapping of the Coulomb problem into
the oscillator: A comment on the papers by Bateman et
al. [Am. J. Phys. {\bf 60} (9), 833-836 (1992)], and P. Pradhan
[Am. J. Phys. {\bf 63} (7), 664 (1995)]'', Am. J. Phys. {\bf 64},
1074-1075 (1996).

\item[{23.}] A.C. Chen, ``Coulomb-Kepler problem and the harmonic
oscillator'', Am. J. Phys. {\bf 55}, 250-252 (1987).
\end{enumerate}

\end{document}